\documentstyle[12pt]{article}
\def\beq{\begin{equation}}
\def\eeq{\end{equation}}
\def\nn{\nonumber}
\def\A{&\hskip-.9em}
\def\mev{\,{\rm MeV}}
\def\csb{\Lambda_{\rm CSB}}
\def\qcd{\Lambda_{\rm QCD}}
\def\DEl{\Delta_H}
\def\del#1{\Delta_{#1}}
\def\prediction{\del{B}=(m_c/m_b)\del{D}}
\def\dm#1#2{m_{{#1}^*_{#2}}-m_{{#1}_{#2}}}
\def\dmp#1#2{(\dm{#1}{#2})}
\def\tr#1{{\rm Tr}[#1]} \def\ih#1{{i#1\over2}}
\def\ch{\bar{H}_a^i}
\def\hai{H^a_i}
\def\hbi{H^b_i}
\def\haj{H^a_j}
\def\hbj{H^b_j}
\def\vmu{v_\mu} \def\dmu{\partial^\mu} \def\xidag{\xi^\dagger}
\def\sigup{\sigma^{\mu\nu}} \def\sigdn{\sigma_{\mu\nu}}
\def\f{f_\pi}
\def\mQ{m_Q}
\def\mQij{(m_Q^{-1})^j_i}
\def\mQij2{(m_Q^{-2})^j_i}
\def\ds{m_{H_s}-m_{H_{ns}}}
\def\dBs{m_{B_s}-m_{B_{ns}}}
\def\oone{\lambda \tr{\ch\sigup\haj\sigdn}(m_Q^{-1})^j_i}
\def\otwo{\lambda'\tr{\ch\sigup\hbj\sigdn}(m_Q^{-1})^j_i{(m_q)^a_b\over\csb}}

\def\othree{\lambda''\tr{\ch\sigup\hbj\sigdn}(m_Q^{-2})^j_i
{(m^2_q)^a_b\over \csb}}
\def\Oone{{\cal O}_1}
\def\Otwo{{\cal O}_2}
\def\Othree{{\cal O}_3}
\begin{document}
\title{Heavy Meson Hyperfine Splittings:\\ A Puzzle for Heavy Quark
Chiral Perturbation Theory\thanks{This work is supported
in part by funds provided by the U.S.
Department of Energy (DOE) under contract \#DE-AC02-76ER03069 and in
part by the Texas National Research Laboratory Commission
under grant \#RGFY92C6.\hfill\break
\vskip 0.05cm
\noindent $^{\dagger}$National
Science Foundation Young Investigator Award.\hfill\break
Alfred P.~Sloan
Foundation Research Fellowship.\hfill\break
Department of Energy Outstanding Junior
Investigator Award.\hfill\break
}}
\renewcommand{\baselinestretch}{1.0}
\author{Lisa Randall$^{\dagger}$
and Eric Sather\\
Massachusetts Institute of Technology\\
Cambridge, MA 02139\\
}
\date{}
\maketitle
\renewcommand{\baselinestretch}{1.2}
\abstract{
We show that there is a large discrepancy between the expected
light
flavor dependence of the  heavy
pseudoscalar--vector mass splittings
and the measured values. We demonstrate that
the one--loop calculation is unreliable. Moreover, agreement
with experiment requires the
leading dependence on SU(3) symmetry breaking to be
nearly cancelled,
so that the heavy quark mass
dependence is unknown and
the expected dependence on the light quark mass is not realized.}
\vfill
November 1992 \hfill  MIT-CTP\#2166
\thispagestyle{empty}
\newpage

\section{Introduction}

Much attention has been devoted to the heavy quark chiral
effective theory \cite{wise}.  The idea is to incorporate
both heavy quark and chiral symmetry into an effective theory
which can describe heavy meson interactions with low momentum
pions. In addition to tree level predictions of the form factors,
such a theory yields a way to predict or at least estimate
the size of SU(3) violating effects.

However, SU(3) violating predictions have not yet been
experimentally tested.  In this paper, we use the heavy
quark chiral effective theory to estimate and calculate
the SU(3) violating parameter
\beq
\del{H}\equiv\dmp{H}{s}-\dmp{H}{ns}
\eeq
where $ns$ stands for nonstrange.
We argue that the estimate one obtains based on a naive expansion in
SU(3) violation is a significant
overestimate. This means that at this point there is no experimental
evidence that an expansion in the strange quark mass
works for heavy quark systems.

  We explicitly calculate
the leading log contribution to verify
the existence of large contributions  to $\del{H}$ which
disagree with the measurements.
We show furthermore that the subleading term
(in powers of $m_s$) contributes as large an amount as the leading
term.
This demonstrates that the procedure of retaining only the one--loop
contribution in chiral perturbation theory is inconsistent.
But the  result that the expansion in SU(3) violation
has not worked   contradicts
what would be naively expected from any model which incorporates
SU(3)
violating light quark masses.

Rosner and Wise \cite{rw}\ recently considered this same parameter,
$\del{H}$.
They enumerated the operators which
are responsible for distinguishing  the various heavy meson masses.
They left the coefficients arbitrary and  fit to existing
data on heavy quark masses.  They then used their assumed dependence
on heavy quark and SU(3) violating parameters to predict
$\prediction$. They concluded that the photons which are emitted
in $B_s^*$ decay and $B^*$ decay should have energies which agree to
within an MeV.

In their paper they observed that the operator which contributes
to $\del{H}$ has a small coefficient.  We claim that this small
coefficient indicates the operator analysis has failed, so one
does not in fact know the leading dependence on heavy quark
parameters.
Because the near cancellation of $\del{D}$ could involve
higher order terms in the $1/m_c$ expansion,
the prediction of $\del{B}$ is not reliable.
%

We reach two conclusions. First, calculations including only
one--loop contributions in the heavy quark chiral effective theory
are not reliable, since the tree level contribution should be
comparable.  Second, there is an interesting physical puzzle as to
why heavy quark chiral perturbation theory does not give the correct
result, even at the order of magnitude level.

In this letter, we first describe the experimental situation,
and review the operator analysis of ref. \cite{rw}. We
estimate the result that  would be expected
on dimensional grounds in the  heavy
quark chiral lagrangian.  The following section
contains a one--loop calculation in which we obtain
a chiral log term in accordance with our estimate.
We discuss possible implications of this result.

\section{Experimental Situation and Operator Analysis}

Much is known about the heavy pseudoscalar-vector meson mass
splittings \cite{e1,e2,e3}:
\begin{eqnarray}
m_{D^{*+}}-m_{D^+}&=&140.64\pm0.08\pm 0.06\mev \\
\dm{D}{s}         &=&141.5\pm 1.9\mev \\
\dm{B}{} &=& 45.4 \pm 1.0\mev\ {\rm (or)}\ 46.2 \pm 0.3 \pm 0.8\mev\\
\dm{B}{s}         &=&47.0\pm 2.6\mev
\end{eqnarray}
It should be observed that the values in the first two lines
and the last two lines are very similar.  The differences
$\del{H}$ are only a few
percent of the SU(3)--symmetric splittings:
\begin{eqnarray}
\del{D}&=&0.9\pm1.9\mev\nn\\
\del{B}&=&1.2\pm2.7\mev.
\end{eqnarray}
The extremely small sizes of these differences are particularly
surprising when compared to what one expects on the basis of a
simple operator analysis, as we now show.

Consider in the chiral heavy quark theory the operators
which contribute to the spin splittings in eqs.\ (1--4) at leading order
in the light quark mass matrix, $m_q$, and the inverse heavy quark
mass matrix, $m_Q^{-1}$.  If the operator which contributes to the
SU(3) symmetric splittings is
\beq
\Oone=\oone,
\eeq
one would expect the operator which contributes to the
chiral symmetry breaking {\it differences}, $\del{D}$ and $\del{B}$,
to be
\beq
\Otwo=\otwo,
\eeq
where the scale of chiral suppression is set by naive application
of the chiral lagrangian dimensional factors. Here $\hai$ is the
field of a heavy meson containing a heavy quark of flavor $i$ and a
light antiquark of flavor $a$. (Note the
value of the light quark masses should be the same as those
taken from fitting the pions, kaons, and nucleons.  There
is an arbitrary strong interaction constant relating the heavy
meson mass splitting to these quark masses, so we cannot
reliably extract the values without a more comprehensive fit.)

However, if this were the case, one would expect the spin splitting
in the $D_s$ system to differ from that in the $D^+$ system by about
$0.15 \cdot 141\mev \approx 20\mev$. As can be seen from the
measured value of $\del{D}$, this is a significant overestimate, off
by about an order of magnitude.  (By comparing the spin-splitting of
heavy mesons containing an strange quark with the splitting for
heavy mesons containing a $down$ quark (so that $\del{H}$ measures
$V$-spin breaking), we avoid the small electromagnetic splittings.
Electromagnetic interactions contribute to  $(D^{*0}-D^0)-(D^{*+}-D^+)$ which
has been measured  to be $1.48\pm0.09\pm0.05$\ \cite{e1}
and is in accordance with expectations.) Notice that
this difficulty in understanding the dependence of the heavy meson
spin splittings  on the strange quark mass is in contrast to our
experience with light baryons, whose spin splittings are well
described by a    nonrelativistic quark model where the  magnetic
moments of the quarks are inversely proportional to their masses.

One might assume that the  operator responsible for SU(3) dependence
of the spin--dependent mass splittings is suppressed. We show in the
next section that this would then be inconsistent at the one loop
level. We calculate an explicit contribution to the splitting that
agrees well with the above estimate.  In fact, if one kept only the
chiral log correction, as has been done in various papers on chiral
heavy quark theory, one calculates a difference of  about $15\mev$
for the $D$ system.  With subleading terms included, the predicted
value is even larger.

\section{One-Loop Calculation of $\del{H}$}

The method of calculation is by now standard. We assume the heavy meson
effective lagrangian, given by
\begin{eqnarray}
{\cal L}=&-&i\tr{\ch\vmu\dmu\hai}
   +\ih{ }\tr{\ch\hbi}\vmu
(\xidag\dmu\xi+\xi\dmu\xidag)^a_b
\nn\\
 &+&\ih{g}\tr{\ch\hbi\gamma_\mu\gamma_5}(\xidag\dmu\xi-\xi\dmu\xidag)^
a_b
     -\rho\tr{\ch\hbi}(m_q)^a_b
\nn\\
&+&\oone.
\end{eqnarray}
Here $\xi=\exp(i\pi^\alpha T^\alpha/\f)$ and $v$ is the heavy quark velocity.
We have kept the leading chiral symmetry breaking contribution
to the masses of the heavy mesons explicit in the lagrangian.
In practice, we use the experimental value of the
strange-nonstrange heavy meson
mass difference to fix $\rho$,
which automatically incorporates the  correct leading
contribution to the  mass differences
due to SU(3) symmetry breaking. We have also explicitly included
the leading heavy quark symmetry breaking operator, $\Oone$, which
is suppressed by $1/m_Q$.
This appears as an explicit vertex in the calculation.

We calculate two types of diagrams. In the first,  Figure 1, we
insert the spin dependent operator $\Oone$, and have a
pseudogoldstone boson emitted and absorbed through the axial
coupling. In the second, Figure 2, we calculate the SU(3) wave
function renormalization   which also contributes at the same order,
given the existing spin splitting, and contributes exactly $-3$
times the amount of the graph in Figure~1 (to all orders in $\ds$).
The result is
\def\pre#1{{g^2\over16\pi^2{#1}}}   \def\hlf#1{{{#1}\over2}}
\def\msqr#1{m_{#1}^2} \def\fsqr#1{f_{#1}^2}
\def\msqrlg#1{ {\msqr{#1}\over\fsqr{#1}} \ln{\msqr{#1}\over\csb^2} }
\begin{eqnarray}
{\del{H}^0\over\dm{H}{}}\A=\A{\dmp{H}{s}-\dmp{H}{ns}\over\dm{H}{}}=\nn\\
\pre{}\biggl\{\A-\A\hlf{3}\msqrlg{\pi}+\msqrlg{K}
+\hlf{1}\msqrlg{\eta}\biggr\}.
\end{eqnarray}
Applying this result to the $D$ system yields a mass splitting of
$\del{D}^0=-15\mev$ if we take $g^2=0.5$.
In the $B$ system $\del{B}^0=-5.1\mev$.

To check the consistency of the calculation as an expansion in the
chiral symmetry breaking parameter, $m_s$, we also calculate
the $m_s^{3/2}$ contribution which results from a linearly divergent
loop integral. It is
\beq
{\del{H}^1\over\dm{H}{}}=\pre{\fsqr{K}}\{-6\pi m_K(\ds)\}, \\
\eeq
where $\ds$ is the heavy-quark symmetric strange-nonstrange heavy
meson  mass splitting which is
fit to be $99.5 \pm 0.6 \mev$ for the $D$ system \cite{e2}
and found to be 80--130$\mev$ for the $B$ system \cite{cusb90}.
This contribution is given by the difference
of an $H_s$ meson
self-energy graph with an intermediate $H$ meson and an
$H$ meson self-energy graph with an intermediate $H_s$
meson. The strange-nonstrange mass splitting contributes with
opposite signs in these two graphs so that in the difference of the
two graphs these terms add constructively.
In the contributions that are zeroth order (i.e., $\del{H}^0$)
and second order in $\ds$
there are cancellations between the $H_s$ and $H$ meson
self-energy graphs.  So while the term second order in
$\ds$ turns out to be
 negligible, the term linear in $\ds$ is larger than the
zeroth order term, $\del{H}^0$,
contributing
$\del{D}^1=-32\mev$ in the $D$ system and $\del{B}^1=-11\mev$
in the $B$ system (using the central value for $\dBs$).
Note that these contributions reinforce the
$\del{H}^0$ contributions found above and give $\del{D}=-47\mev$ and
$\del{B}=-16\mev$.  There are also extra finite pieces zeroth order
in $\ds$ that are quadratic in the pseudogoldstone masses.  These are also
comparable in size to the log terms in $\del{H}^0$.
The large size of each of  the non-log terms shows that retaining
only log terms is not a reasonable approximation.

This calculation demonstrates that the parameter $\DEl$ will
not scale linearly with the strange quark mass, since terms
proportional to $m_s$ and $m_s^{(3/2)}$ were of comparable
importance.  Although not manifest in this calculation,
which was only done to order $1/m_Q$, we demonstrate that
straightforward estimates of the size of terms which are higher
order in inverse powers of the heavy quark mass are also
far from negligible.

This can be seen by an operator analysis analogous to that in
the second section, but including higher order operators.  For
example, the operator
\beq
\Othree=\othree
\eeq
should contribute about $1.5 \mev$ to  $\del{D}$, which is still larger
than the measured value.

In fact, a one--loop estimate of the contribution to $\del{D}$ at
order $1/m_c^2$ is even larger.
%
%
To study the $1/m_c^2$ terms,
we could insert into the heavy quark line
in Figure 1 or 2 the two-derivative piece
of the heavy meson kinetic term,
${1\over2}\tr{\ch\partial^2\haj}(m_Q^{-1})^i_j$, or we could simply insert
the spin-spitting operator, $\Oone$, a second time.  We would expect
contributions to $\del{H}/\dmp{H}{}$ of order $m_K^3/(\mQ\csb^2)$
and $m_K\dmp{H}{}/\csb^2$ respectively.
This is actually larger than the contribution from
$\Othree$ above.  For the $D$ system, these
terms each contribute of order $10\mev$ to $\del{D}$.

Clearly we
cannot account for the very small size of $\del{D}$ if we  truncate
its expansion in powers of $1/m_c$ at the first term, proportional
to $1/m_c$. All we know is that there is a conspiracy between
a large number of terms generated at tree and loop level, all
of which individually would generate a large contribution to $\del{c}$,
but whose sum is small.
Therefore, we cannot assume  $\prediction$. Furthermore,
because we do not know the role of the higher order terms in the
cancellation which produces a small value of $\del{D}$, we
cannot necessarily conclude that $\del{B}$ is  small, although
preliminary measurements do give a small value.

\section{Discussion}

This result is clearly disturbing. The picture of the heavy meson
based on leading SU(3) and heavy quark symmetric physics broken
at order $\qcd/m_Q$ and $m_s/\csb$ does not
predict the correct size of $\del{H}$, assuming the existing
data is correct. This behavior is peculiar from any viewpoint,
independent of the heavy quark effective theory,
since
one would naively expect a fairly large effect due to the
fact that the magnetic moment of the strange constituent
quark is less than that of the non--strange counterpart by a
significant
amount. For example,
using a quark model, Close  in 1979 \cite{close} predicted
  $\dm{D}{s}\approx{2\over3}\dmp{D}{}$.
We also evaluated using the bag model
the change in color magnetostatic energy for a strange quark mass of
$200\mev$
(a small value from the standpoint of the bag model, since the
net mass contributed to the meson is then only about $100\mev$).
The strange heavy meson magnetostatic energy
in this model  was also about 2/3 of the corresponding
value for a nonstrange heavy meson.  However, it is
interesting to note that an estimate by Godfrey and Isgur \cite{gi}
predicted $\del{D}\approx -10\mev$.

This disagreement between the expected and measured dependence
on the strange quark mass is clearly a puzzle. From the
standpoint of the expansion in chiral symmetry breaking and the
inverse of the heavy quark mass, it would be attributable
to a cancellation among tree and loop contributions.
The role of higher order terms in $1/m_Q$
will be put to the test when the photon
energies for the transitions $B^*_s\to B_s\gamma$ and $B^{*0}\to
B^0\gamma$
are more accurately determined. A photon energy larger than an MeV
will demonstrate the importance of higher order terms in
the inverse of the heavy quark mass; a small value
would however be inconclusive, since it could arise from
a cancellation between tree and loop contributions
to $\del{D}$ at any given order in $1/m_c$.

If there is an
accidental cancellation in the one measured
quantity computed with heavy quark chiral perturbation theory,
it is possible this can happen elsewhere. It would
be difficult to determine which predictions are reliable.
One might hope  this calculation is somehow distinct
from others which have been done.  However this is difficult
to reconcile with the fact that wave function renormalization
alone would in itself generate a large effect.

The small value of $\del{H}$ might signal
something fundamental about the heavy meson,
indicating that the operator analysis is not the best description.
 For example, the
authors of ref. \cite{rw} suggested that
the SU(3) breaking
 strong hyperfine splitting due to the change in chromomagnetic
moments of the light quarks is cancelled by a change
in the wave function of the heavy meson. It would be
interesting to test other predictions of the heavy quark theory.
For example, the transition magnetic moment of the heavy meson \cite{8a}, the
flavor dependence of $f_H$ and
$B_H$
calculated in ref. \cite{grinetal} or of the Isgur--Wise function
\cite{jensav} would all be interesting
measurements, if they can be done.
These measurements would answer the following
questions:
1) Do the one--loop calculations give the correct result?
2) Are there other parameters whose values do not have the
expected dependence on light quark mass? and 3) If there
is a model in which the wave function at zero interquark separation
cancels the dependence on the quark mass of the magnetic moment,
does it give correct predictions for these other quantities?
These would not only settle the issue of whether the discrepancy
between the values of $\DEl$ obtained from  the expected
chiral expansion and from experiment  was purely accidental,
but could also test  the flavor
dependence of the wave function of the heavy meson.

We conclude that the discrepancy between the expected and measured
values of $\DEl$ is a very interesting puzzle. If the data in
the $D$ and $B$ systems is correct, we might have an interesting
probe of heavy mesons at hand.
On the other hand, if there is simply an accidental cancellation
between large terms, it would be worth investigating if
this happens in other measurable parameters as well.
It would be useful to supplement
this measurement with other measurements of SU(3) violating effects
to test the validity of possible proposed forms of light quark flavor
dependence.

 \section*{Acknowledgements}
 We are grateful to Mike Dugan, Howard Georgi, Mitch Golden,
 Bob Jaffe, Shmuel Nussinov,      Jim Olness, and Mark Wise
 for discussions and to Jon Rosner for discussing the results of ref.
\cite{rw}.

\def\thefiglist#1{\section*{Figure Captions\markboth
 {FIGURE CAPTIONS}{FIGURE CAPTIONS}}\list
 {Figure \arabic{enumi}.}
 {\settowidth\labelwidth{Figure #1.}\leftmargin\labelwidth
 \advance\leftmargin\labelsep
 \usecounter{enumi}\parsep 0pt \itemsep 0pt plus2pt}
 \def\newblock{\hskip .11em plus .33em minus -.07em}
 \sloppy}
\let\endthefiglist=\endlist
\begin{thefiglist}{9}
\item Diagram with an insertion of the spin dependent operator,
$\Oone$. The dotted line is a pseudogoldstone boson
(which can be strange or nonstrange) and the solid line
is the heavy meson (which can be strange or nonstrange, spin one or spin zero).
\item  Wave function renormalization. Notation same as for Figure~1.
\end{thefiglist}


\begin{thebibliography}{9}
  \bibitem{wise} M. Wise, Phys. Rev. {\bf D45} (1992) 3021.
  \bibitem{rw} J. Rosner and M. Wise, CALT--68--1897, EFU--92--40.
  \bibitem{e1}CLEO Collaboration, D. Bortoletto {\it et. al.},
  ``Isospin Mass Splittings from Precision Measurements of $D^*$-$D$
Mass
  Differences," Cornell University Report No. CLNS 92/1152, submitted
  to Phys. Rev. Lett.
  \bibitem{e2} Particle Data Group, K.~Hikasa {\it et. al.},
Phys.
  Rev. {\bf D 45} (1992).
  \bibitem{e3}CUSB Collaboration. J. Lee-Franzini {\it et. al.},
Phys. Rev.
  Lett. {\bf 67}, 1692 (1991).
  \bibitem{cusb90} CUSB Collaboration. J. Lee-Franzini {\it et. al.},
Phys. Rev.
  Lett. {\bf 65}, 2947 (1990).
  \bibitem{close} F. Close, \lq \lq An Introduction to Quarks and
Partons",
  Academic Press (1979).
  \bibitem{gi} S. Godfrey and N. Isgur, Phys. Rev. {\bf D 32} (1985), 189.
  \bibitem {grinetal} B. Grinstein, E. Jenkins, A. Manohar, M.
Savage,
  and M. Wise, Nucl. Phys. {\bf B380} (1992) 369.
  \bibitem{8a} J. Amundson, C. G. Boyd, E. Jenkins, M. Luke, A.
Manohar,
  J.~Rosner, M. Savage, M. Wise, UCSD/PTH 92--31, hep--ph/9209241.
  \bibitem{jensav} E. Jenkins and M. Savage Phys. Lett. {\rm B281}
(1992) 331.
  \end{thebibliography}
\end{document}